\documentclass[aps,pre,twocolumn,groupedaddress,showpacs]{revtex4-1}

\usepackage{natbib}
\usepackage{graphics}
\usepackage{amsfonts}
\usepackage{amssymb}
\usepackage{amsmath}

\begin{document}

\title{Equation of state of hard oblate ellipsoids by replica exchange Monte Carlo}

\author{G. Odriozola}
\email{godriozo@imp.mx} 

\author{F. de J. Guevara-Rodr\'{\i}guez}
\email{fguevara@imp.mx} 

\affiliation{Programa de Ingenier\'{\i}a
Molecular, Instituto Mexicano del Petr\'{o}leo, Eje Central
L\'{a}zaro C\'ardenas 152, 07730, M\'{e}xico, Distrito Federal,
M\'{e}xico.}

\date{\today}

\begin{abstract}
We implemented the replica exchange Monte Carlo technique to produce the equation of state of hard 1:5 aspect-ratio oblate ellipsoids for a wide density range. For this purpose, we considered the analytical approximation of the overlap distance given by Bern and Pechukas and the exact numerical solution given by Perram and Wertheim. For both cases we capture the expected isotropic-nematic transition at low densities and a nematic-crystal transition at larger densities. For the exact case, these transitions occur at the volume fraction $0.341$, and in the interval $0.584-0.605$, respectively.
\end{abstract}


\maketitle

Anisotropic molecules and colloids show a strong tendency to assemble into complex structures which range from liquid-crystals~\cite{Care05,Wilson05} to empty liquids~\cite{Ruzicka11a,Ruzicka11b}. There are basically two approaches to model these systems: site by site or atomistic, and coarse graining~\cite{Care05,Wilson05}. In the first approach detail is usually gained at the expenses of a larger computational cost. In the second, many sites are grouped into entities which, in many cases, do not show a spherical symmetry. In such cases, ellipsoidal pair potentials have been widely used~\cite{Perram84,Perram85,Paramonov05}. Among them, probably the most popular is the soft Gay-Berne interaction~\cite{Gay81,Berardi09,Ghoufi11,Luckhurst10,Lee10}, which is based on the hard core overlap distance introduced by Berne and Pechukas (BP)~\cite{Berne72}. 

An ellipsoid can be handled to match the shape of many molecules and colloids. For instance, the hard core of laponites, a clay with a well-defined disk-like shape, can be approached by oblates having a 1:25 aspect-ratio~\cite{Odriozola04,Mossa07,Patti11,Ruzicka11a,Ruzicka11b}. Then, charge or Yukawa sites can be added to match their electrostatic properties. This surely leads to a considerable decrease of computational cost since otherwise the clay body must be modeled with a large number of beads to mimic its hard core~\cite{Odriozola04}. For this purpose, however, it is desirable to know the exact shape of the model hard core, as well as its corresponding volume and surface. Unfortunately, the BP hard potential does not provide a defined shape and volume. This may explain the subsequent efforts to efficiently solve the exact ellipsoidal hard core interaction~\cite{Perram84,Perram85}. In this regard, Paramonov and Yalirakia~\cite{Paramonov05}, based on the work of Perram and Wertheim (PW)~\cite{Perram84,Perram85}, recently presented a numerical algorithm for determining the directional contact distance of two generic ellipsoids, which can be used to detect overlaps. 


The BP hard potential~\cite{Berne72} is analytical, mathematically simple, easy to implement, fast to compute, and it can be used to study the condensed phase of prolate or oblate particles via simulations~\cite{Ghoufi11,Luckhurst10,Berardi09,Luckhurst93}. In this approach molecules are represented with an uniaxial ellipsoidal Gaussian and their interaction is then related to their overlap integral. This way, a coarse-grained potential is built which successfully captures the anisotropic nature of the entities. The expression for the distance between the geometric centers of the ellipsoids when the particles are at contact, $\sigma_{BP}$, is given by $\sigma_{BP} = \sigma_{\perp} ( 1 - \frac{1}{2} \chi [ A^{(+)} + A^{(-)} ] )^{-1/2}$, where $A^{(\pm)} = (\hat{\mathbf{r}}_{ji} \cdot \hat{\mathbf{u}}_{i} \pm \hat{\mathbf{r}}_{ji} \cdot \hat{\mathbf{u}}_{j} )^{2} / (1 \pm \chi \hat{\mathbf{u}}_{i} \cdot \hat{\mathbf{u}}_{j})$, $\hat{\mathbf{u}}_{i}$ and $\hat{\mathbf{u}}_{j}$ are the versors along the axial axis of each particle, and $\hat{\mathbf{r}}$ is the versor along the line joining the geometric centers. Here, $\chi=(\sigma_{\|}^{2}-\sigma_{\bot}^{2}) / (\sigma_{\|}^{2}+\sigma_{\bot}^{2})$ is the anisotropy parameter, where $\sigma_{\|}$ and $\sigma_{\bot}$ are the parallel and perpendicular diameters with respect to the axial axis, respectively.


On the other hand, an exact solution for determining whether two ellipsoids overlap or not can be numerically obtained. An ellipsoidal surface centered at $\mathbf{r}_{i}$ and oriented according to $\hat{\mathbf{u}}_{i}$ is given by the following quadratic form $\mathcal{A}_{i}( \mathbf{r}_e ) = (\mathbf{r}_e - \mathbf{r}_{i})^{t} \cdot \mathbb{A}_{i} \cdot (\mathbf{r}_e - \mathbf{r}_{i})$, where $\mathbf{r}_e$ is a point at the surface $\mathcal{A}_{i}( \mathbf{r}_e ) = 1$, $\mathbb{A}_{i} = \mathbb{U}^{t}( \hat{\mathbf{u}}_{i} ) \cdot \mathbb{D}^{-2} \cdot \mathbb{U}( \hat{\mathbf{u}}_{i} )$, 
$\mathbb{U}( \hat{\mathbf{u}}_{i} )$ is the rotation matrix, $\mathbb{U}^{t}$ is its transpose, and  
$\mathbb{D} = \frac{1}{2} \sum_i \sigma_i \hat{e}_i \otimes \hat{e}_i$ ($\hat{e}_i$ is a versor along the particle principal axis). Let's consider two arbitrarily oriented ellipsoids $i$ and $j$ at contact at point $\mathbf{r}_{c}$. The vector normal to the $i$ surface at $\mathbf{r}_{c}$ is $\mathbf{n}_{i}( \mathbf{r}_{c} ) = \mathbb{A}_{i} \cdot (\mathbf{r}_{c} - \mathbf{r}_{i})$. A similar relation can be written for the vector normal to the $j$ surface $\mathbf{n}_{j}$. Since the tangent plane is common for both ellipsoids, the normal versors $\hat{\mathbf{n}}_{i}=\mathbf{n}_{i}/|\mathbf{n}_{i}|$ and $\hat{\mathbf{n}}_{j}=\mathbf{n}_{j}/|\mathbf{n}_{j}|$ fulfill                                                                                                                                                                                                                                                                                                                                                                                                                                                                                                                          
$\hat{\mathbf{n}}_{i}( \mathbf{r}_{c} ) + \hat{\mathbf{n}}_{j}( \mathbf{r}_{c} ) = \mathbf{0}$. This condition was originally employed by PW for developing an algorithm to numerically determine the point $\mathbf{r}_c$~\cite{Perram84,Perram85}. In their work the Elliptic Contact Function (ECF) is introduced. This function contains the above given information and allows to determine the distance of the closest approach. Later, the ECF procedure was reviewed by Paramonov and Yaliraki, who contributed with a clear geometrical interpretation of the PW approach~\cite{Paramonov05}. In particular, the expression for the function that connects the particles centers trough the geometrical place where the vectors $\nabla \mathcal{A}_{i}( \mathbf{x}_{c} )$ and $\nabla \mathcal{A}_{j}( \mathbf{x}_{c} )$ are antiparallel is given by~\cite{Paramonov05} $\mathbf{x}_{c}( \lambda ) = ( \lambda \mathbb{A}_{i} + ( 1 - \lambda ) \mathbb{A}_{j} )^{-1} \cdot ( \lambda \mathbb{A}_{i} \cdot \mathbf{r}_{i} + ( 1 - \lambda ) \mathbb{A}_{j} \cdot \mathbf{r}_{j} ) $, where $\lambda \in [0,1]$ is a scalar parameter. Note that for $\lambda=1$ and 0 the geometric centers of the ellipsoids $i$ and $j$ are obtained, respectively. The contact point $\mathbf{r}_c$ lies on this trajectory and corresponds to a unique value of $\lambda$, $\lambda_{c}$, which fulfills $\lambda_{c} \in (0,1)$. Furthermore, $\mathcal{A}_{i} (\mathbf{r}_c ) = \mathcal{A}_{j} ( \mathbf{r}_c )$, with $\mathbf{r}_c= \mathbf{x}_{c}( \lambda_{c})$.

With the above expressions it is easy to implement an iterative procedure to yield $\mathbf{r}_c$ with the desired precision. In particular, we start by evaluating $\Delta(\lambda)=\mathcal{A}_{i} (\mathbf{x}_c(\lambda)) - \mathcal{A}_{j} (\mathbf{x}_c(\lambda))$ for $\lambda=0.5$. A positive $\Delta(\lambda)$ means $1>\lambda_c>\lambda$ and so, we increase $\lambda$ in such a way to reduce in half the interval. Conversely, a negative $\Delta(\lambda)$ means $0<\lambda_c<\lambda$ and we accordingly decrease $\lambda$. This way the interval is reduced as $1/2^n$, being $n$ the number of iterations. Approximately 20 iterations yield an error of $\Delta$ smaller that $1 \times 10^{-6}$. Note that the involved operations are products and summations which translate into a relatively fast computation. 

Additionally, the contact parameter $\lambda_{c}$ is the extreme value of the linear combination of the quadratic forms $\mathcal{A}_{i}( \mathbf{x}_c )$ and $\mathcal{A}_{j}( \mathbf{x}_c )$, i.~e.,
$ \mathcal{S}( \lambda ) = \lambda \mathcal{A}_{i} ( \mathbf{x}_{c}( \lambda ) ) + ( 1 - \lambda ) \mathcal{A}_{j} ( \mathbf{x}_{c}( \lambda ) )$ and $0 \leq \mathcal{S}( \lambda ) \leq \mathcal{S}( \lambda_{c} )$~\cite{Paramonov05}. This property can also be used to numerically determine $\lambda_c$. The contact parameter defines the PW contact distance, $\sigma_{PW} = r / \sqrt{ \mathcal{S}( \lambda_{c} ) } $~\cite{Perram84,Perram85,Paramonov05}. Consequently, ellipsoids having their geometric centers separated at a distance $r<\sigma_{PW}$ overlap whereas they do not for $r>\sigma_{PW}$. In particular, the BP analytical expression for the contact distance corresponds to $ \sigma_{BP} = r / \sqrt{\mathcal{S}(1/2)}$~\cite{Paramonov05}. This expression makes clear that $\sigma_{BP} \geq \sigma_{PW}$. 


Even though the analytical expressions for determining whether or not two ellipsoids overlap are relatively fast to compute, sampling from crowded systems is always a difficult task~\cite{Odriozola11}. Thus, we implemented the replica exchange Monte Carlo methodology, which is well proven to assist the systems to reach equilibrium at difficult (high density / low temperature) conditions~\cite{swendsen,hukushima96,Lyubartsev92,Marinari92,Berardi09,Odriozola09}. Since we are dealing with hard ellipsoids we must perform the replica expansion in pressure. Hence, the partition function of the extended ensemble is given by~\cite{Okabe01,Odriozola09} $ Q_{\rm extended}=\prod_{i=1}^{n_r} Q_{N T P_i}$, where $Q_{NTP_i}$ is the partition function of the isobaric-isothermal ensemble of the system at pressure $P_i$, temperature $T$, and particle number $N$. This extended ensemble is sampled by combining standard $NTP_i$ simulations on each replica and swap moves at the replica level. These swap moves are performed by means of the following acceptance probability~\cite{Odriozola09} $P_{\rm acc}\!=\! \min(1,\exp[\beta(P_i-P_j)(V_i-V_j)])$, where $V_i-V_j$ is the volume difference between replicas $i$ and $j$. More details on this method are given in refs.~\cite{Odriozola09,Odriozola11}.

Simulations are started by randomly placing and orienting the ellipsoids (avoiding overlaps), so that the initial volume fraction is $\varphi= v_{e} \rho = 0.2$, where $\rho$ is the number density, $v_{e}=4 \pi \sigma_{\|} \sigma_{\bot}^2/3$ is the ellipsoid volume (for both studied models), $\sigma_{\bot}=5\sigma_{\|}$, and $\sigma_{\|}$ is taken as the length unit. We first perform $2 \times 10^{13}$ trial moves at the desired state points, during which we observe the replicas reaching a stationary state. We then sample by performing additional $2 \times 10^{13}$ trials. This work is performed by considering $N=100$ and $n_r=64$.


\begin{figure}
\resizebox{0.47\textwidth}{!}{\includegraphics{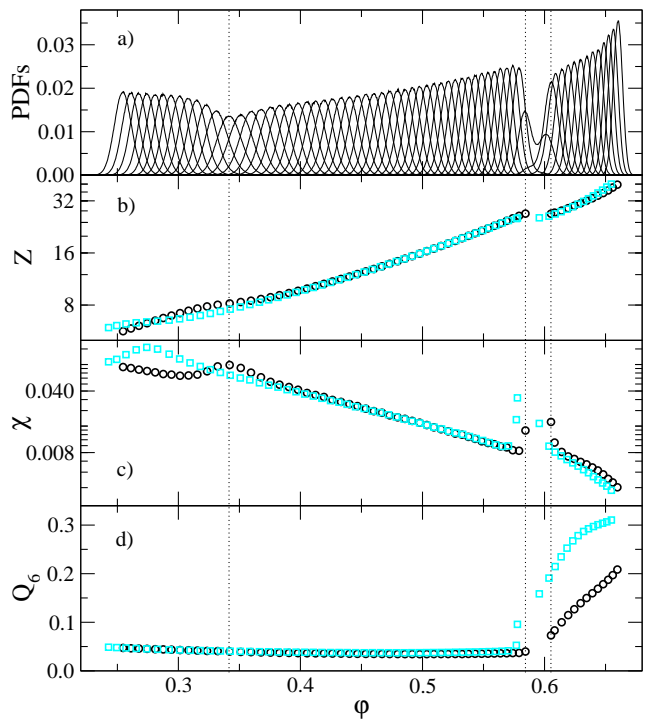}}
\caption{\label{Fig1} a) Probability distribution functions (PDFs) of volume fraction fluctuations for each of the $n_r=64$ pressure values, and for the exact hard 1:5 oblate ellipsoidal model. b) Equations of state, $Z(\varphi)$. c) Isothermal compressibilities, $\chi(\varphi)$, obtained from density fluctuations. d) Order parameters, $Q_6(\varphi)$. For panels b), c), and d), dark circles and light squares correspond to the exact PW and the BP analytical solution for the overlap distance, respectively. Vertical dotted lines highlight the PW transitions. }
\end{figure}

The exact hard 1:5 oblate ellipsoidal model can be studied since, on the one hand, the exact iterative procedure to solve the $\sigma_{PW}$ overlap distance is relatively fast, and on the other, the numerical solution is computed only for those cases where $r<\sigma_{BP}$ (otherwise ellipsoids do not overlap). The probability distribution functions (PDFs) obtained for this system are shown in Fig.~\ref{Fig1} a). There are 64 curves corresponding to each fixed pressure, which increases from left to right. The general trend of the PDFs is to get narrower and higher with increasing pressure evidencing a decrease of the isothermal compressibility $\chi= \delta \rho / \delta (\beta P)$. However, at $\varphi \approxeq 0.34$ and at $\varphi \approxeq 0.60$ this trend is disrupted. Here the PDFs turn wider, shorter, and distorted, pointing out phase transitions. In particular, at $\varphi \approxeq 0.60$ PDFs are bimodal, which is the typical behavior of a first order transition (a discontinuity of the pressure as a function of the density). Hence, from this plot one expects three different phases each one corresponding to different density regions.

Panels b) and c) of Fig.~\ref{Fig1} are built from the PDFs shown in panel a). There, the dimensionless pressure $Z=\beta P/\rho$ is plotted (dark circles) as well as the isothermal compressibility $\chi$, both as a function of the most frequent $\varphi$. The $\chi$ values are obtained by the density fluctuations, i.~e., by $\chi=N(<\rho^2>-<\rho>^2)/<\rho>^2$, which should equal $\chi= \delta \rho / \delta (\beta P)$ according to the fluctuation-dissipation theorem. Finally, panel d) shows (dark circles) the order parameter $Q_6$, as defined in refs.~\cite{Steinhardt96,Rintoul96b,Odriozola09}, implemented for the geometric centers of the ellipsoids. The isotropic-nematic (liquid-liquid crystal) phase transition is observed as a plateau of $Z(\varphi)$ and no discontinuity is detected. However, this transition is clearly pointed out by the large jump of $\chi(\varphi)$, which peaks at $\varphi = \varphi_{i-n} \approxeq 0.341$. Additionally, $Q_6(\varphi)$ shows no sign of change at this density, indicating the expected absence of positional order. The second transition occurs in the interval $\varphi_f=0.584<\varphi<\varphi_s=0.605$ and corresponds to $Z \approxeq 27.0$ (for $N\rightarrow \infty$ we expect a shift of $\varphi_f$, $\varphi_s$, and $Z$ to larger values~\cite{Odriozola09}). It is characterized by a discontinuity of $Z(\varphi)$, and a jump of $\chi(\varphi)$ and $Q_6(\varphi)$. This $Q_6(\varphi)$ jump is indicative of the appearance of positional order of the ellipsoids' geometric centers. It should be noted that the fluid-crystal transition interval for hard ellipsoids is well shifted to the right as compared to the hard spheres transition ($\varphi_f=0.492<\varphi<\varphi_s=0.545$)~\cite{Wilding00,Noya08,Odriozola09}. Something similar happens to the maximum hard ellipsoid packing fraction \cite{Donev04}. A fit of the form $Z \sim (\varphi_d-\varphi)^{-1}$ to the high pressure branch of $Z(\varphi)$ leads to a divergence at $\varphi_d=0.768$. This value far exceeds the largest packing fraction of hard spheres, $\varphi=0.74$, and is close to the maximum density of hard ellipsoids reported by Donev et.~al., $0.77$ \cite{Donev04}.


Panels b), c), and d) of Fig.~\ref{Fig1} also include the results obtained for the BP analytical approximation as light squares (these results are 3 times faster to obtain than those corresponding to the exact PW solution). As it can be seen, for $0.37 \lesssim \varphi \lesssim 0.57$, the three plots, $Z(\varphi)$, $\chi(\varphi)$, and $Q_6(\varphi)$ show a very good match between BP and PW. However, both BP transitions are shifted to lower densities, which is a direct consequence of the systematical overestimation of the contact distance, $\sigma_{BP} \geq \sigma_{PW}$. Since BP clearly overestimates the T-shape configuration the equations of states quantitatively disagree for $\varphi \lesssim 0.37$, when T-shape configurations are relevant. This disagreement is clearly seen in panels b) and c) of Fig.~\ref{Fig1}. Conversely, the T-shape configurations are practically absent in the nematic phase and thus the BP expression matches the exact PW solution. At very high pressures the systematic overestimation of $\sigma_{BP}$ produces a less dense crystal branch, an underestimation of $\chi(\varphi)$, and a more structured crystal (see $Q_6(\varphi)$). 

\begin{figure}
\resizebox{0.48\textwidth}{!}{\includegraphics{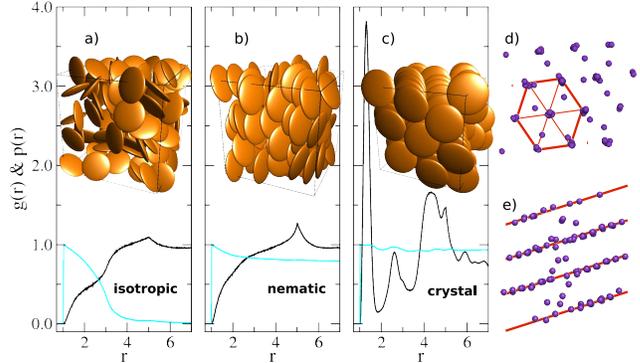}}
\caption{\label{Fig2} Radial distribution functions, $g(r)$, (dark lines) and radial order parameters, $p(r)$, (light lines) for a) the lowest, b) an intermediate, and c) the highest pressure. The insets are corresponding snapshots. d) Front view and, e), side view of the corresponding geometric centers of the inset at c). Results are obtained for the exact $\sigma_{PW}$.}
\end{figure}

The structure of the different phases is analyzed in Fig.~\ref{Fig2} where the radial distribution function and order parameter, $p(r)=<1/2(3(\hat{\mathbf{u}}_{i} \cdot \hat{\mathbf{u}}_{j})^2-1)>$, are shown together with the corresponding snapshots for the highest (panel c)), the lowest a), and an intermediate pressure b). The corresponding phases are crystal, isotropic, and nematic, respectively. The isotropic phase is characterized by a slowly increasing $g(r)$ which peaks at $r\approxeq \sigma_{\bot}$. The $p(r)$ shows the typical peak at $r = \sigma_{\|}$, which practically vanishes at $r\approxeq \sigma_{\bot}$. Thus, there is not a long range alignment of the entities for $\varphi<\varphi_{i-n}$. The corresponding snapshot clearly shows this fact (inset of panel a)). Conversely, the fluid structure for $\varphi_{i-n}<\varphi<\varphi_f$ does show a long range alignment of the ellipsoids, as indicated by the large value of $p(r)$ for all distances, $r$ (panel b)). However, the ellipsoids' geometric centers still present a fluid-like structure, as pointed out by the $g(r)$ function. Finally, for $\varphi>\varphi_s$ the $g(r)$ grows several peaks pointing out a highly structured phase (panel c)). The first and larger one appears at $r\approxeq 1.2$, and corresponds to the touching and stacked ellipsoids shown in the inset. The second, at $r\approxeq 2.4$, is also due to the stacked arrangement and corresponds to two ellipsoids separated by a third one and sandwiched by them. The wide peak at $r\approxeq 4.2$ corresponds to side-to-side configurations of particles belonging to different stacks. It is located at $r < \sigma_{\bot}$ since the particles of the adjacent stacks are partially sandwiched. The $p(r)$ function also shows for this case the expected long range alignment. More details on the obtained crystal structure are seen in panels d) and e). Panel d) highlights the obtained hexagonal arrangement of the particles' geometric centers (in correspondence to the snapshot inserted in panel c)). Additionally, panel e) shows the side view of the stacks' axes, which align defining planes.  


In summary, this work provides the equation of state of hard 1:5 aspect-ratio oblate ellipsoids by means of replica exchange simulations for a moderate system size ($N=100$). This is done by considering the BP analytical expression and the exact numerical solution for determining overlaps. We observed a good overall agreement between BP and the exact numerical solution, though deviations appear at low concentrations, where T-shape configurations are frequently present in the system structure. For both cases, an isotropic-nematic and a nematic-crystal transitions are captured, occurring for the exact case at $\varphi_{i-n}=0.341$ and $Z=8.2$, and in the interval $\varphi_f-\varphi_s=0.584-0.605$ and $Z=27.0$, respectively.   

The authors thank projects Nos.~Y.00116 and Y.00119 SENER-CONACyT for financial support.


\begin{thebibliography}{10}

\bibitem{Care05}
C.~M. Care and D.~J. Cleaver, Rep. Prog. Phys. {\bf 68},  2665  (2005).

\bibitem{Wilson05}
M.~R. Wilson, Int. Rev. Phys. Chem. {\bf 24},  421  (2005).

\bibitem{Ruzicka11a}
B. Ruzicka and E. Zaccarelli, Soft Matter {\bf 7},  1268  (2011).

\bibitem{Ruzicka11b}
B. Ruzicka {\it et~al.}, Nature Materials {\bf 10},  56  (2011).

\bibitem{Perram84}
J.~W. Perram, M.~S. Wertheim, J.~L. Lebowitz, and G.~O. Williams, Chem. Phys.
  Lett. {\bf 105},  277  (1984).

\bibitem{Perram85}
J.~W. Perram and M.~S. Wertheim, J. Chem. Phys. {\bf 58},  409  (1985).

\bibitem{Paramonov05}
L. Paramonov and S.~N. Yaliraki, J. Chem. Phys. {\bf 123},  194111  (2005).

\bibitem{Gay81}
J.~G. Gay and B.~J. Berne, J. Chem. Phys. {\bf 74},  3316  (1981).

\bibitem{Berardi09}
R. Berardi, C. Zannoni, J.~S. Lintuvuori, and M.~R. Wilson, J. Chem. Phys. {\bf
  131},  174107  (2009).

\bibitem{Ghoufi11}
A. Ghoufi, D. Morineau, R. Lefort, and P. Malfreyt, J. Chem. Phys. {\bf 134},
  034116  (2011).

\bibitem{Luckhurst10}
G.~R. Luckhurst and K. Satoh, J. Chem. Phys. {\bf 132},  184903  (2010).

\bibitem{Lee10}
C.~K. Lee, C.~C. Hua, and S.~A. Chen, J. Chem. Phys. {\bf 133},  064902
  (2010).

\bibitem{Berne72}
B.~J. Berne and P. Pechukas, J. Chem. Phys. {\bf 56},  4213  (1972).

\bibitem{Odriozola04}
G. Odriozola, M. Romero-Bastida, and F.~de J. Guevara-Rodriguez, Phys. Rev. E.
  {\bf 70},  021405  (2004).

\bibitem{Mossa07}
S. Mossa, C. De~Michele, and F. Sciortino, J. Chem. Phys. {\bf 126},  014905
  (2007).

\bibitem{Patti11}
A. Patti, S. Belli, R. van Roij, and M. Dijkstra, Soft Matter {\bf 7},  3533
  (2011).

\bibitem{Luckhurst93}
G.~R. Luckhurst and P.~S.~J. Simmonds, Mol. Phys. {\bf 80},  233  (1993).

\bibitem{Odriozola11}
G. Odriozola and L. Berthier, J. Chem. Phys. {\bf 134},  054504  (2011).

\bibitem{swendsen}
R.~H. Swendsen and J.-S. Wang, Phys. Rev. Lett. {\bf 57},  2607  (1986).

\bibitem{hukushima96}
K. Hukushima and K. Nemoto, J. Phys. Soc. Jpn. {\bf 65},  1604  (1996).

\bibitem{Lyubartsev92}
A.~P. Lyubartsev, A.~A. Martsinovski, S.~V. Shevkunov, and P.~N.
  Vorontsov-Velyaminov, J. Chem. Phys. {\bf 96},  1776  (1992).

\bibitem{Marinari92}
E. Marinari and G. Parisi, Europhys. Lett. {\bf 19},  451  (1992).

\bibitem{Odriozola09}
G. Odriozola, J. Chem. Phys. {\bf 131},  144107  (2009).

\bibitem{Okabe01}
T. Okabe, M. Kawata, Y. Okamoto, and M. Mikami, Chem. Phys. Lett. {\bf 335},
  435  (2001).

\bibitem{Steinhardt96}
P.~J. Seinhardt, D.~R. Nelson, and M. Ronchetti, Phys. Rev. B. {\bf 28},  784
  (1996).

\bibitem{Rintoul96b}
M.~D. Rintoul and S. Torquato, J. Chem. Phys. {\bf 105},  9258  (1996).

\bibitem{Wilding00}
N.~B. Wilding and A.~D. Bruce, Phys. Rev. Lett. {\bf 79},  3002  (1997).

\bibitem{Noya08}
E.~G. Noya, C. Vega, and E. de~Miguel, J. Chem. Phys. {\bf 128},  154507
  (2008).

\bibitem{Donev04}
A. Donev, F.~H. Stillinger, P.~M. Chaikin, and S. Torquato, Phys. Rev. Lett.
  {\bf 92},  P07015  (2004).

\end{thebibliography}

\end{document}